\documentclass[prl,aps,groupedaddress,showpacs,twocolumn,floatfix,nofootinbib]{revtex4}
\usepackage{amsmath,amsthm,amsfonts,amssymb,graphics,graphicx,epsfig,color,times,natbib,subfigure}
\usepackage{bbm} % for \mathbbm{1}...
\usepackage{bm}

\newcommand{\be}{\begin{equation}}
\newcommand{\ee}{\end{equation}}
\newcommand{\ba}{\begin{eqnarray}}
\newcommand{\ea}{\end{eqnarray}}

\newcommand{\one}{\leavevmode\hbox{\small1\normalsize\kern-.33em1}}

\definecolor{nblue}{rgb}{0.2,0.2,0.7}
\definecolor{ngreen}{rgb}{0.2,0.6,0.2}
\definecolor{nred}{rgb}{0.8,0.2,0.2}
\definecolor{nblack}{rgb}{0,0,0}
\definecolor{mar}{rgb}{0.6,0.1,0.1}

\newcommand{\xfrac}[2]{{#1}/{#2}}
\renewcommand{\section}[1]{{\em #1}.---}

\newcommand{\cu}[1]{\left\{ {#1} \right\}}

\begin{document}

\title{One-sided Device-Independent Quantum Key Distribution: \\ Security, feasibility, and the connection with steering}
\author{Cyril Branciard$^1$, Eric G. Cavalcanti$^2$, Stephen P. Walborn$^3$, Valerio Scarani$^{4,5}$, and Howard M. Wiseman$^{2,5}$}
\affiliation{$^1$ School of Mathematics and Physics, The University of Queensland, St Lucia, QLD 4072, Australia \\
$^2$ Centre for Quantum Dynamics, Griffith University, Brisbane QLD 4111, Australia \\
$^3$ Instituto de F\'{i}sica, Universidade Federal do Rio de Janeiro, Caixa Postal 68528, Rio de Janeiro, RJ 21941-972, Brazil \\
$^4$ Centre for Quantum Technologies and Department of Physics, National University of Singapore, Singapore 117543 \\
$^5$ ARC Centre for Quantum Computation and Communication Technology, Griffith University, Brisbane QLD 4111, Australia}
\date{\today}

\begin{abstract}

We analyze the security and feasibility of a protocol for Quantum Key Distribution (QKD), in a context where only one of the two parties trusts his measurement apparatus. This scenario lies naturally between standard QKD, where both parties trust their measurement apparatuses, and Device-Independent QKD (DI-QKD), where neither does, and can be a natural assumption in some practical situations. We show that the requirements for obtaining secure keys are much easier to meet than for DI-QKD, which opens promising experimental opportunities. We clarify the link between the security of this one-sided DI-QKD scenario and the demonstration of quantum steering, in analogy to the link between DI-QKD and the violation of Bell inequalities.

\end{abstract}

\pacs{03.67.Dd, 03.65.Ud, 03.67.Mn}   
 %03.65.Ud Entanglement and quantum nonlocality
%03.67.Dd Quantum cryptography and communication security
%03.67.Mn Entanglement measures, witnesses, and other characterizations
\maketitle

%\section{Introduction}
Quantum Key Distribution (QKD) allows two parties (Alice and Bob) to establish secret keys at a distance, with security guaranteed by the laws of Quantum Mechanics~\cite{QKD_review}. In standard QKD (S-QKD), the security is typically proved under the assumption that Alice and Bob can trust the physical functioning of their preparation and measurement apparatuses. For instance, standard security proofs for the BB84 protocol~\cite{BB84} assume that Alice sends qubits to Bob, prepared in some eigenstates of the $\sigma_z$ or $\sigma_x$ Pauli operators, and that Bob measures them in one of those two bases. Recent demonstrations of hacking of the devices has shown the importance and weakness of this assumption~\cite{hacking}. Moreover, since QKD is becoming commercially available, Alice and Bob may end up buying their devices from untrusted providers. 

Remarkably, there are ways to guarantee security with fewer assumptions. The minimal set of assumptions is the one used in Device-Independent QKD (DI-QKD)~\cite{DIQKD_MayersYao,DIQKD_PRL}. There, Alice and Bob can certify the security of QKD based only on the observed violation of Bell inequalities~\cite{bell64}: the measurement apparatuses are untrusted black boxes, with a knob supposedly related to the measurement settings. Alice and Bob have only to trust the random number generator with which they vary the positions of the knob, and of course the integrity of their locations. While the qualitative understanding ``Bell violation implies security" is certainly true, the derivation of quantitative security bounds is challenging. The most recent results report on security against the most general attacks (``coherent attacks") under the assumption that previous measurements do not feed any information forward to subsequent ones~\cite{MPA11,HR09} (or, to put it simply: that the devices are memoryless). In addition to these (hopefully temporary) limitations of the theoretical studies, DI-QKD imposes very demanding requirements on practical demonstrations. In particular, the Bell test would need to close the detection loophole~\cite{detLH}, which requires very high detection efficiencies~\cite{DIQKD_NJP}.

Intermediate scenarios between S-QKD and DI-QKD require less trust than the former and will be easier to implement than the latter~\cite{pawlowski_brunner_semiDIQKD}. Imagine for instance that a bank wants to establish secret keys with its clients; the bank could invest a lot of money to establish one trustworthy measurement device, but the clients at the other end of the channel would certainly get cheap (and insecure) detection terminals. This leads us to study {\em one-sided} DI-QKD (1sDI-QKD): we consider an entanglement-based scenario in which Bob's measurement apparatus is trusted, while Alice's is not. In the entanglement-based setup the source is also untrusted, although we will also discuss prepare-and-measure (P\&M) implementations where the source is trusted. We present a security bound against coherent attacks with similar assumptions as in Refs.~\cite{MPA11,HR09}, in particular that the devices are memoryless, as this enables the strongest security analysis presently available. Focusing on practical implementations, we show that the detector efficiencies required for a practical implementation of 1sDI-QKD are much lower than for DI-QKD, making it feasible with existing devices. Before that, let us start by stressing a link with a hierarchy of tests of quantum nonlocality~\cite{steering_PRL}.

\section{QKD and quantum nonlocality}
\begin{figure}%[!h]
\begin{center}
\epsfxsize=7.5cm
\epsfbox{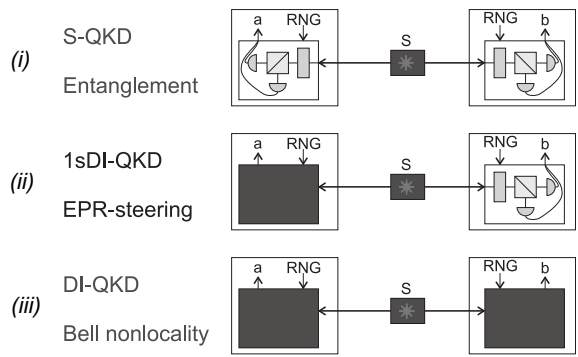}
\caption{Link between the three concepts of quantum nonlocality as classified in~\cite{steering_PRL} and the three scenarios of S-QKD, 1sDI-QKD (this paper) and DI-QKD.
In order to obtain a secret key, {\it (i)} if Alice and Bob trust their measurement devices (transparent boxes), then they must necessarily demonstrate {\it entanglement}; {\it (ii)} if Alice's measurement device is untrusted (black box), while Bob's is trusted, then Alice must demonstrate {\it steering} of Bob's state; {\it (iii)} if both Alice's and Bob's measurement devices are untrusted, then they must demonstrate {\em Bell-nonlocality}. In all cases, Alice and Bob must trust their random number generator (RNG), and the integrity of their location.}
\label{fig_scenarios}
\end{center}
\end{figure}

It is known that no secret key can be extracted in a QKD experiment if the channel between Alice and Bob is entanglement-breaking~\cite{curty_04}. Hence, in order to demonstrate security, one must show that the channel preserves entanglement. The three different assumptions on Alice's and Bob's devices mentioned above correspond naturally to three different criteria for quantum nonlocality~\cite{steering_PRL} (see Figure~\ref{fig_scenarios}): S-QKD or DI-QKD require the observed correlations to violate a separability criterion or a Bell inequality respectively; 1sDI-QKD requires the correlations to violate an \textit{EPR-steering inequality} as defined in~\cite{CJWR09}. That is, if one imagines that Bob's system has a definite (albeit unknown to him) quantum state, the protocol must prove that Alice, by her choice of measurement, can affect this state. This sort of nonlocality, first discussed by Einstein, Podolsky and Rosen \cite{EPR35}, was called `steering' by Schr\"odinger~\cite{Sch35b}.
In Ref.~\cite{steering_PRL} these concepts of nonlocality arose from considering entanglement verification with untrusted {\em parties}. However, even if Alice and Bob trust each other, as in QKD, they may not trust their devices, which is an analogous situation. From this perspective, our scenario of 1sDI-QKD can thus be seen as a practical application of the concept of {\it quantum steering}.

\section{A 1sDI-QKD protocol}
We consider the following 1sDI version of the BBM92 entanglement-based protocol~\cite{bbm92}: Alice and Bob receive some (typically, photonic) quantum systems from an external source. Alice can choose between two binary measurements, $A_1$ and $A_2$; since she does not trust her measurement device, she treats it as a black box with two possible settings, yielding each time one of two possible outputs.
Bob, on the other hand, trusts his device to make projective measurements $B_1$ or $B_2$ in some qubit subspace, typically corresponding to the operators $\sigma_z$ and $\sigma_x$ respectively. After publicly announcing which measurements they chose for each system, Alice and Bob will try to extract a secret key from the conclusive results of the measurements $A_1$ and $B_1$; as explained below, the results of measurements $A_2$ and $B_2$ will allow them to estimate Eve's information.

Alice and Bob might not always detect the photons sent by the source, because of losses or inefficient detectors. Since Bob trusts his detectors, he trusts that Eve cannot control his detections. Also, Eve cannot get any useful information from Bob's (null) result if the photons going to him are lost or if she keeps them. Cases where Bob gets ambiguous results (e.g. double clicks) can be dealt with using the techniques of Ref.~\cite{moroder_squashing}--see Supplemental Material \cite{SM} for details. Hence, we can safely consider only the cases where Bob gets detections. On the other hand, since Alice's measurement device is untrusted, Eve could control whether her detectors click depending on the state she receives and on her choice of measurement setting. We can therefore not simply discard Alice's no-detection events. In case her detectors don't click, she records a bit value of her choice as the result of her measurement, keeps track of the fact that her detectors did not click, and tells Bob (so that they can later post-select the raw key on Alice's detections); Eve has access to that information.

We denote by ${\boldsymbol A}_i$ and ${\boldsymbol B}_i$ the strings of classical bits Alice and Bob get from measurements $A_i$ and $B_i$ (and where Bob got a detection, as discussed above). Among the bits of ${\boldsymbol A}_1$, some correspond to actual detections by Alice, and some, corresponding to non-detections, were simply chosen by Alice herself. Everyone knows which ones are which. The detected bits form a string ${\boldsymbol A}_1^{\rm ps}$ (they'll be {\it post-selected} by Alice and Bob), while those that were not actually detected form a string ${\boldsymbol A}_1^{\rm dis}$ (they'll be {\it discarded} for the key extraction), so that ${\boldsymbol A}_1 = ({\boldsymbol A}_1^{\rm ps},{\boldsymbol A}_1^{\rm dis})$. Bob's corresponding bit strings are ${\boldsymbol B}_1^{\rm ps}$ and ${\boldsymbol B}_1^{\rm dis}$, resp., so that ${\boldsymbol B}_1 = ({\boldsymbol B}_1^{\rm ps},{\boldsymbol B}_1^{\rm dis})$. We denote by $N$ the length of the strings ${\boldsymbol A}_1$ and ${\boldsymbol B}_1$, and by $n$ that of the strings ${\boldsymbol A}_1^{\rm ps}$ and ${\boldsymbol B}_1^{\rm ps}$.

\section{Security proof and key rate}
Recently, Tomamichel and Renner~\cite{TR11}, together also with Lim and Gisin~\cite{TomaLimGiRenner} have developed an approach to QKD based on an uncertainty relation for smooth entropies, which enables one to prove security against coherent attacks in precisely this 1sDI-QKD scenario; note however that one also needs (as in~\cite{MPA11,HR09}) the assumption that the devices are memoryless~\cite{fehr_renner}. To prove the security of our protocol in realistic implementations, we extend their analysis by considering imperfect detection efficiencies~\cite{endnote:2}.

From the $n$-bit strings ${\boldsymbol A}_1^{\rm ps}$ and ${\boldsymbol B}_1^{\rm ps}$, on which Eve may have some (possibly quantum) information $E$, Alice and Bob can extract, through classical error correction and privacy amplification (from Bob to Alice), a secret key of length \cite{renes_renner}
\ba \label{rr}
\ell & \approx & H_{\mathrm{min}}^\epsilon({\boldsymbol B}_1^{\rm ps}|E) - n \, h(Q_1^{\rm ps}) \,.
\ea
Here $H_{\mathrm{min}}^\epsilon({\boldsymbol B}_1^{\rm ps}|E)$ denotes the {\it smooth min-entropy}~\cite{smooth_ent} of ${\boldsymbol B}_1^{\rm ps}$, conditioned on quantum side information $E$; $h$ is the binary entropy function: $h(Q) \equiv - Q \log_2Q - (1-Q) \log_2(1-Q)$; and $Q_1^{\rm ps}$ is the bit error rate between ${\boldsymbol A}_1^{\rm ps}$ and ${\boldsymbol B}_1^{\rm ps}$.

To bound $H_{\mathrm{min}}^\epsilon({\boldsymbol B}_1^{\rm ps}|E)$, we will use the uncertainty relation introduced in~\cite{TR11}, which bounds Eve's information on $B_1$ given Alice's information on the incompatible observable $B_2$. However, we need to use the full strings ${\boldsymbol B}_1, {\boldsymbol B}_2$, as post-selection may lead to an apparent violation of the uncertainty relation.
Using the chain rule~\cite{renes_renner} and the data-processing inequality~\cite{TomamichelColbeckRenner} for smooth min-entropies, we first bound Eve's information on ${\boldsymbol B}_1^{\rm ps}$ relative to her information on ${\boldsymbol B}_1$:
\ba
H_{\mathrm{min}}^\epsilon({\boldsymbol B}_1|E) &=& H_{\mathrm{min}}^\epsilon({\boldsymbol B}_1^{\rm ps},{\boldsymbol B}_1^{\rm dis}|E) \\
& \leq & H_{\mathrm{min}}^\epsilon({\boldsymbol B}_1^{\rm ps}|{\boldsymbol B}_1^{\rm dis}E) + \log_2|{\boldsymbol B}_1^{\rm dis}| \\
& \leq & H_{\mathrm{min}}^\epsilon({\boldsymbol B}_1^{\rm ps}|E) + N-n \,. \label{ps_bound}
\ea

Now, consider a hypothetical run of the protocol where the bits of ${\boldsymbol A}_1$ and ${\boldsymbol B}_1$ would be measured in the second basis; we denote by ${\boldsymbol A}_2$ and ${\boldsymbol B}_2$ the corresponding hypothetical strings. From the generalized uncertainty relation of~\cite{TR11}, one has 
\ba
H_{\mathrm{min}}^\epsilon({\boldsymbol B}_1|E) & \geq & q N - H_{\mathrm{max}}^\epsilon({\boldsymbol B}_2|{\boldsymbol A}_2) \,. 
\label{TR_UR}
\ea
where $q$ is a measure of how distinct Bob's two measurements are; for orthogonal qubit measurements, $q=1$. 
Here, $H_{\mathrm{max}}^\epsilon({\boldsymbol B}_2|{\boldsymbol A}_2)$ is the {\it smooth max-entropy}~\cite{smooth_ent} of ${\boldsymbol B}_2$, conditioned on ${\boldsymbol A}_2$. It satisfies the following~\cite{TomaLimGiRenner}
\ba
H_{\mathrm{max}}^\epsilon({\boldsymbol B}_2|{\boldsymbol A}_2) \ \lesssim \ N \, h(Q_2) \,,
\label{H_h_2}
\ea
where $Q_2$ is the bit error rate between ${\boldsymbol A}_2$ and ${\boldsymbol B}_2$.
Now, since the choice of basis was made randomly, $Q_2$ is the same as the bit error rate observed---without post-selection---when the second basis was actually chosen (no matter how rarely) by both Alice and Bob.

Substituting \eqref{ps_bound}, \eqref{TR_UR} and \eqref{H_h_2} in Eq.~(\ref{rr}), we obtain
\ba
\ell & \gtrsim & n \, [1 - h(Q_1^{\rm ps})] - N\,[h(Q_2) + 1 - q] \,. \label{eq_l_1sDI_approx}
\ea
In the asymptotic limit of infinite key lengths, the above approximate inequality becomes exact~\cite{footnote_finite_key_effects,TomaLimGiRenner}.
The fraction $n/N $ of photons which Alice detects, given that Bob detected one, will be denoted $\eta_A$. This allows us to write the {\it secret key rate} $r \equiv \ell/N$ (the number of secret bits obtained per photon detected by Bob, measured in the first basis), as
\ba
r & \geq & \eta_A [1 - h(Q_1^{\rm ps})] - h(Q_2) -(1-q)\, . \label{eq_r_1sDI}
\ea

\section{Relation to EPR-steering}
As we recalled before, the secret key rate above can only be positive if Alice and Bob can check that they share entanglement. In our 1sDI scenario, this amounts to demonstrating quantum steering. Hence, the inequality  $\eta_A [1 - h(Q_1^{\rm ps})] - h(Q_2) -(1-q) \leq 0$ can be understood as an EPR-steering inequality~\cite{CJWR09}. In the Supplemental Material \cite{SM}, we give a more direct proof of this claim, starting from the so-called {\it Local Hidden State model}~\cite{steering_PRL}.

\section{Experimental prospects}
We now turn to the feasibility analysis. Consider a typical experimental setup, where a source sends maximally entangled 2-qubit states to Alice and Bob, through a depolarizing channel with visibility $V$, and where, as in the BBM92 protocol,  $A_1 = B_1 = \sigma_z$ and $A_2 = B_2 = \sigma_x$, with Alice's detection efficiency being $\eta_A$ as above. (We emphasize that this is simply a model for Alice's measurements, which are implemented in a black box.)

\begin{figure}%[!h]
\begin{center}
\epsfxsize=8cm
\epsfbox{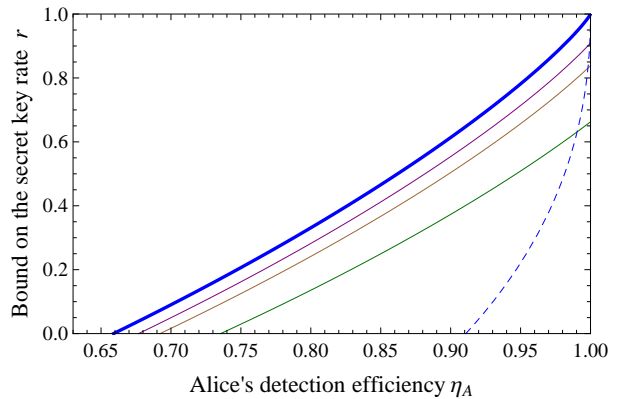}
\caption{Solid curves: bounds (\ref{eq_r_1sDI}) on the secret key rate $r$ in a typical implementation of 1sDI-QKD, as a function of Alice's detection efficiency $\eta_A$, for visibilities $V = 1, 0.99, 0.98, 0.95$ (from top to bottom) and $q=1$. Dashed curve: for comparison, bounds (for $V=1$) for DI-QKD, obtained by adapting the security analysis of Ref.~\cite{MPA11}, when Bob has the same detection efficiency $\eta_B = \eta_A$ as Alice (see Supplemental Material \cite{SM}).}
\label{fig_r_eta_V}
\end{center}
\end{figure}

The secret key rate that Alice and Bob can extract is then bounded by (\ref{eq_r_1sDI}), with $q=1$ and
\ba
Q_1^{\rm ps} = \xfrac{(1-V)}{2} \, , \qquad Q_2 = \xfrac{(1-\eta_A V)}{2} \, . \nonumber
\ea
Figure~\ref{fig_r_eta_V} shows the values of the bound~(\ref{eq_r_1sDI}) as a function of $\eta_A$, and for different values of $V$. For a perfect visibility $V = 1$, one gets a positive secret key rate for all $\eta_{A} > 65.9 \%$.

This detection probability threshold is much lower than those required for DI-QKD. For instance, if Alice and Bob have the same detection efficiency $\eta$, then they require $\eta >94.6\%$ for the protocol studied in~\cite{MPA11}, when they extract their key from the non-post-selected data. If the key is extracted from the post-selected data (as we considered here for 1sDI-QKD), the threshold remains quite high, $\eta >91.1 \%$ (see Figure~\ref{fig_r_eta_V} and Supplemental Material \cite{SM}).
The much lower efficiency threshold for 1sDI-QKD compared with DI-QKD is related to the fact that it is much easier to close the detection loophole in a steering experiment~\cite{steering_GU,steering_Vienna,steering_UQ} than in a Bell test, for which there are no photonic detection-loophole free demonstrations to date. Heralding efficiencies of $\sim 62 \%$ have recently been reported~\cite{steering_UQ}, in an experiment demonstrating detection-loophole-free quantum steering; our 1sDI-QKD protocol could be demonstrated with a very similar (but slightly improved) experimental setup.

Note also that in the 1sDI-QKD case, the losses between the source and Bob's lab do not affect the security of the protocol; they only decrease the key rate proportionally to the decrease of Bob's detection rate (as long as the noise in Bob's detectors does not become prominent). Hence, long distances can in principle be reached if the source stays close to Alice; this is in contrast to the fully DI-QKD case, where the limit on the detection efficiencies imposes a limit on the allowed distance between Alice and Bob (although some proposals have been suggested to overcome this problem~\cite{DIQKD_heralded_qubit,curty_moroder,pitkanen}).

\section{Comparison of different scenarios}
Key rate bounds for entanglement-based QKD also apply to P\&M schemes, as long as the preparation device can be trusted to produce a certain {\it average} state independent of Alice's (or Bob's, as the case may be) choice of preparation basis (e.g. the completely mixed state, regardless of whether  $\sigma_x$ or $\sigma_z$ is the chosen basis). A preparation device with this property can be envisaged as a trusted  entanglement source situated in Alice's (or Bob's) laboratory, with a trusted channel between it and the local detector. In this picture it still makes sense to consider the case where the local ``hypothetical'' detector is untrusted, as this is equivalent to saying that we cannot trust the preparation device to prepare the desired state. Here the efficiency (which we will denote as $\eta^*$) of the hypothetical detector models the probability that the preparation device registers to the sender which of the two states (in the chosen basis)  was sent. For a well-functioning device this can be close to unity; even if the preparation does use a probabilistic photon-pair source and a detector, the sender can generate many pairs within the time window for each system, and switch out the system (one photon, ideally) only when its preparation is heralded by the detection of the other. A greater experimental challenge is the loss of the heralded photon (within the sender's lab or {\em en route}), which must be factored into the receiver's efficiency.

Considering all non-trivial permutations of device trustworthiness, there are eight P\&M scenarios, and four genuine entanglement-based scenarios, whose security can be analysed using the methods of Refs.~\cite{TR11} or \cite{MPA11}. We remove mirror-image scenarios by keeping only the version which is better (or equally good) under the assumption of the privacy amplification being from Bob to Alice, as shown in Table~\ref{bounds_table}.
Note that P\&M by Bob with a fully trusted preparation device (row 3) does not improve the threshold efficiency required for Alice's untrusted detector, as compared to steering-based 1sDI-QKD with an untrusted entanglement source (row 6). 

\begin{table} 
\begin{tabular}{|c|ccccccccc|c|c|}
\hline 
Based on & AD && C && S && C && BD & Key rate bound & Eff. Thresh. \tabularnewline
\hline 
\hline 
 P\&M & T && T && T &$\vline\vline$& U &$\vline\vline$& T & $r_0(Q_1^{\rm ps},Q_2^{\rm ps})$ & none \tabularnewline
\hline 
 P\&M &U && T && T &$\vline\vline$& U &$\vline\vline$& T & $r_1(Q_1^{\rm ps},Q_2)$ & $\eta_A^* > 65.9\%$  \tabularnewline
\hline 
 P\&M &U &$\vline\vline$& U &$\vline\vline$& T && T && T & $r_1(Q_1^{\rm ps},Q_2)$ & $\eta_A > 65.9\%$  \tabularnewline 
 \hline
 P\&M &U &$\vline\vline$& U &$\vline\vline$& T && T && U & $r_2(Q_1^{\rm ps},S)$ &  $\eta_A >  83.3\%$ 
\tabularnewline \hline
\hline 
 Entang. & T &$\vline\vline$& U && U && U &$\vline\vline$& T & $r_0(Q_1^{\rm ps},Q_2^{\rm ps})$ & none \tabularnewline
\hline 
Steering & U &$\vline\vline$& U && U && U &$\vline\vline$& T & $r_1(Q_1^{\rm ps},Q_2)$ & $\eta_A > 65.9\%$ \tabularnewline
\hline 
Bell &U &$\vline\vline$& U && U && U &$\vline\vline$& U & $r_2(Q_1^{\rm ps},S)$& $\eta > 91.1\%$ \tabularnewline
\hline 
\end{tabular}
\caption{Best known bounds on secret key rates for QKD (secure against coherent attacks) with privacy amplification from Bob to Alice and memoryless devices, both P\&M and entanglement-based. The second column tells which of the components---Alice's detectors (AD), the source (S), Bob's detector (BD) and each of the channels (C) between the source and the detectors---are trusted (T) or untrusted (U). Thick vertical lines in each row separate Alice's lab, the channel open to Eve, and Bob's lab. In the P\&M cases, the detector plus source in a lab is a formal model for a preparation device; the efficiency $\eta^*$ in these cases models the probability that the preparation device registers which state is prepared, and would typically be close to unity (for details see text).
In column three, the bounds on key rates (here, per {\em photon pair} produced by the source, i.e.~per {\em preparation event} in the P\&M cases) are given in terms of the functions $r_0(Q_1^{\rm ps},Q_2^{\rm ps}) \equiv  \eta_B\eta_A  [1 - h(Q_1^{\rm ps}) - h(Q_2^{\rm ps})]$ from S-QKD, $r_1(Q_1^{\rm ps},Q_2) \equiv  \eta_B\{\eta_A [1 - h(Q_1^{\rm ps})] - h(Q_2)\}$ from Eq.~(\ref{eq_r_1sDI}), and $r_2(Q_1^{\rm ps},S) \equiv \eta_A\eta_B [1-h(Q_1^{\rm ps})] - \log_2\big[1+\sqrt{2-(S/2)^2}\big]$ (see Supplemental Material \cite{SM}). Here $S$ is the value of the CHSH polynomial~\cite{CHSH}, while $Q_1$ and $Q_2$ are bit error rates. The superscript ${\rm ps}$ means post-selection on coincident detections; $Q_2$ in $r_1(Q_1^{\rm ps},Q_2)$ is post-selected on Bob's detections, but not on Alice's; $S$ must be estimated from the whole non-postselected data.
The Efficiency Thresholds (column four) are calculated with everything else perfect. In row 4, we assumed $\eta^*_B \to 1$ (the threshold we quote is therefore a lower bound on the thresholds for $\eta^*_B < 1$), while in row 7, $\eta_A = \eta_B = \eta$.}
\label{bounds_table}
\end{table}

\section{Conclusion} We have introduced the scenario of 1sDI-QKD and analyzed its security against coherent attacks in a practical situation where losses are taken into account. Our analysis shows that the assumptions of 1sDI-QKD allow one to significantly lower the necessary detection efficiencies compared to fully DI-QKD, and that the requirements for obtaining secure key rates in an experiment are within the range of current technology.

We have also stressed that 1sDI-QKD requires the violation of an EPR-steering inequality, in analogy with the requirement of violation of a Bell inequality for security of DI-QKD. The relation between the QKD hierarchy and the nonlocality hierarchy introduces some open questions: (i) It has recently been shown that steering can be demonstrated with arbitrarily low efficiencies~\cite{steering_GU}. Can one find (and prove the security of) 1sDI-QKD protocols that would also tolerate arbitrarily low efficiencies? (ii) With 2 measurement settings per party, steering can be demonstrated for efficiencies $\eta_{A}>50 \%$~\cite{steering_GU,steering_UQ}. There is a large gap between this and the threshold of $\sim 66 \%$ for our 1sDI-QKD protocol. The same situation occurs for fully DI-QKD, where there is also a gap between the threshold of $\eta > 82.8 \%$ for a violation of the CHSH inequality~\cite{CHSH} and that for the security of DI-QKD. How small can these be made in general? Another topic for further research is to extend our results to finite keys, for instance along the lines of Ref.~\cite{TomaLimGiRenner}.

\section{Acknowledgements}
This work was supported by the Australian Research Council Programs CE110001027 and DP0984863, the National Research Foundation and the Ministry of Education, Singapore, the Brazilian agencies CNPq, FAPERJ, and the INCT-Informa\c{c}\~ao Qu\^antica.

\section{Note added}
While finishing writing this manuscript, we became aware of a related work~\cite{ma_lutkenhaus} where similar bounds on the detection efficiency thresholds are derived.

\onecolumngrid
\clearpage

\renewcommand{\theequation}{S\arabic{equation}}
\setcounter{equation}{0}

\begin{center}
{\bf \large Supplemental Material}
\end{center}

\bigskip
\bigskip

\twocolumngrid

\section{Practical detectors and multiple counts}
In practical implementations with photons, Bob's detector comprises two photon counters, corresponding to his two possible outputs. Because of dark counts or imperfect state preparation (i.e. the presence of multi-photon states) there will be rare occasions where both counters click. In such cases, our protocol requires Bob to output a random result. This is a standard way to treat double clicks in the context of QKD~\cite{lutkenhaus_SM}. When Bob implements the $\sigma_z$ or $\sigma_x$ measurements, his detection scheme has a ``squashing property'' that should allow one to extend the present theoretical security proof for perfect qubit states (on Bob's side) to imperfect implementations~\cite{moroder_squashing_SM}.

\bigskip

\section{Direct proof that ``$r \leq 0$'' is a steering inequality}
We obtained in Eq.~(8) in the main text %~(\ref{eq_r_1sDI}) 
a bound on the secret key rate of our 1sDI-QKD protocol. As we argued, a positive secret key rate can be obtained only if Alice and Bob can demonstrate steering. That is, if they {\it violate} the inequality ``r.h.s of Eq.~(8) %(\ref{eq_r_1sDI}) 
$\leq 0$'', or more explicitly
\ba
\eta h(Q_1^{\rm ps}) + h(Q_2) + (1{-}\eta) & \geq & q \, , \label{eq_steering_ineq}
\ea
then they can conclude that they have demonstrated steering: Eq.~(\ref{eq_steering_ineq}) thus defines an {\it EPR-steering inequality}~\cite{CJWR09_SM}.

\medskip

Here we show in a different way that~(\ref{eq_steering_ineq}) is indeed an EPR-steering inequality, starting more directly from the {\it Local Hidden State model}~\cite{CJWR09_SM}---a local model whereby the statistics observed by Alice and Bob can be decomposed as~\cite{steering_PRL_SM}
\ba
P(\tilde{a}_i,b_j)=\sum_\lambda P(\lambda) P(\tilde{a}_i|\lambda) P_Q(b_j|\lambda), \label{LHS_model}
\ea
where $\tilde{a}_i = (a_i,s)$, with $a_i = \pm 1$ and $b_j = \pm 1$ denoting the experimental outcomes corresponding to measurements $A_i$, $B_j$, and $s$ denoting whether Alice gets a detection ($s=1$) or not ($s=0$); recall that in case of no detection, Alice's result $a_i$ is chosen to be a pre-established bit in our 1sDI-QKD protocol and will contribute to the bit string ${\boldsymbol A}_i^{\rm dis}$. The subscript $Q$ indicates that Bob's statistics are determined by quantum mechanics: $P_Q(b_j|\lambda) = {\rm Tr}[E_j \rho_\lambda]$, where $\rho_\lambda$ is some quantum state (in our case, a qubit state) and $\cu{E_j}$ are the positive operators summing to unity which describe his measurements.

The variable $\lambda$ fully specifies the quantum state received by Bob. As a consequence, Alice can have no more information on average about Bob's results than what one could know through $\lambda$. In terms of conditional entropies, this implies that
\ba
H(B_i|\tilde{A}_i) & \geq & H(B_i|\Lambda) \,, \label{bound_LHS}
\ea
where $H(B_i|\tilde{A}_i) = \sum_{\tilde{a}_i} P(\tilde{a}_i) H(B_i|\tilde{a}_i)$  and $H(B_i|\Lambda) = \sum_{\lambda} P(\lambda) H(B_i|\lambda)$ are the average entropies of Bob's measurement $B_i$ conditioned on the knowledge of the result of $\tilde{A}_i$ and of $\lambda$, respectively, with $H(B_i|\tilde{a}_i) = - \sum_{b_i} P(b_i|\tilde{a}_i) \log_2 P(b_i|\tilde{a}_i)$ and $ H(B_i|\lambda) = - \sum_{b_i} P_Q(b_i|\lambda) \log_2 P_Q(b_i|\lambda)$.

Now, Bob's statistics, conditioned on $\lambda$, are by assumption given by a quantum state; they must therefore satisfy all quantum uncertainty relations, and in particular
\ba
H(B_1|\lambda) + H(B_2|\lambda) \geq q,
\ea
where $q$ quantifies how different the measurements $B_1$ and $B_2$ are~\cite{uncert_relation_q_SM}. 
Together with~\eqref{bound_LHS}, we obtain
\ba
H(B_1|\tilde{A}_1) + H(B_2|\tilde{A}_2) \geq q. \label{eq:steer_ineq_H}
\ea

Denoting $A_i^{\rm ps}$ for $A_i$ when $s_i=1$ and $A_i^{\rm dis}$ for $A_i$ when $s_i=0$, and assuming 
that $P(s_i=1) = \eta_A$ independently of $A_i$ (as should be the case if it is a detector efficiency),  we have
\ba
H(B_i|\tilde{A}_i) = \eta_A H(B_i^{\rm ps}|A_i^{\rm ps}) + (1{-}\eta_A) H(B_i^{\rm dis}|A_i^{\rm dis}). \quad \label{eq:decomp_HBA}
\ea

Now, one can easily check (using for instance the concavity of the binary entropy function) that for binary variables $A$ and $B$, one has
\ba
\label{eq:step_4}
H(B|A) \leq h(Q),
\ea
where $Q$ is the bit error rate relative to $A$ and $B$ (the probability that $A \neq B$). 
Applying this to~\eqref{eq:decomp_HBA} for $i=1$, it follows that 
\ba
H(B_1|\tilde{A}_1) & \leq & \eta_A h(Q_1^{\rm ps}) + (1-\eta_A), \label{H_leq_h_1}
\ea
while on the other hand 
\begin{equation}
H(B_2|\tilde{A}_2)  \leq  H(B_2|{A}_2) \leq  h(Q_2). \label{H_leq_h_2}
\end{equation}
Substituting \eqref{H_leq_h_1} and \eqref{H_leq_h_2} in \eqref{eq:steer_ineq_H} we finally obtain \eqref{eq_steering_ineq}, as desired. 

Note that for the purpose of deriving an EPR-steering inequality, there is no need to follow the last steps of the above procedure where we treat and bound $H(B_1| \tilde{A}_1) $ and $H(B_2|\tilde{A}_2)$ differently. Rather we can use Eq.~(\ref{eq:steer_ineq_H}) directly as an 
EPR-steering inequality. Under the usual conditions for QKD as discussed in the main text,  $q=1$ and Eq.~(\ref{H_leq_h_1}) 
becomes a tight inequality, so Eq.~(\ref{eq:steer_ineq_H}) then becomes 
\begin{equation}
\eta_A[ 2 - h(Q_1^{\rm ps}) - h(Q_2^{\rm ps})] \leq 1 .
\end{equation}
For perfect correlations ($Q_1^{\rm ps} = Q_2^{\rm ps} = 0$), this will be violated (indicating EPR-steering) provided that $\eta_A > \frac{1}{2}$. This is the minimum possible efficiency threshold for EPR-steering using two settings~\cite{steering_GU_SM}.

\bigskip

\section{Comparison with DI-QKD}
In the DI-QKD case, the strongest security analysis available today is that of Masanes, Pironio and Ac\'in~\cite{MPA11_SM}, who showed the security of a large class of protocols against coherent attacks (under the technical assumption that the devices are  memoryless). In the simplest version, the protocol they consider involves Alice and Bob measuring in bases that enable a CHSH inequality~\cite{CHSH_SM} to be tested, where one of Bob's bases is that which is used (together with a different one of Alice's bases) to yield the secret shared key. Let the CHSH polynomial have the value $S$ (such that $|S|<2\sqrt{2}$ for quantum correlations, and $|S|>2$ indicates a violation of the CHSH inequality), and let the non-post-selected quantum bit error rate for measurements in the bases that will yield the key be $Q_1$. Then Ref.~\cite{MPA11_SM} shows that the final key rate can be bounded from below by 
\be
r_{\rm MPA} = 1-h(Q_1) - \log_2\big[1+\sqrt{2-(S/2)^2}\big] .\label{rMPA}
\ee
For a visibility $V$ and inefficiencies $\eta_A$ and $\eta_B$, we have a raw shared random key with bit error rate $Q_1 = [1-(1-\eta_A)(1-\eta_B)-V\eta_A\eta_B]/2$, and a CHSH parameter $S = 2\sqrt{2}  V\eta_A \eta_B + 2(1-\eta_A)(1-\eta_B)$. The latter is obtained if Alice and Bob use the same predetermined list of random results whenever they fail to detect a photon.
With $V=1$ and $\eta_A=\eta_B=\eta$, Eq.~(\ref{rMPA}) yields a positive key rate for $\eta > 94.5\%$. In the situation of prepare-and-measure (P\&M), where there is a trusted source in Bob's lab, but his detector is still untrusted, we replace $\eta_B$ by $\eta_B^*$, Bob's preparation efficiency. Taking $\eta_B^*\to 1$, Eq.~(\ref{rMPA}) gives  a lower bound $\eta_A^- > 89.6\%$.

\medskip

We now show that Alice and Bob can improve their  secret key rate by extracting their secret key from the data post-selected on coincidence detections, as we suggested here for the case of 1sDI-QKD. Indeed, one can apply our analysis to that of Ref.~\cite{MPA11_SM} to find a new secret key rate bounded from below by 
\be
r_{2} = \eta_A\eta_B[1-h(Q_1^{\rm ps})] - \log_2\big[1+\sqrt{2-(S/2)^2}\big] .\label{eq:r2}
\ee
Here $S$ is unchanged from above (it must still be estimated from the whole non-postselected data), while $Q_1^{\rm ps}=(1-V)/2$. With $V=1$ and $\eta_A=\eta_B=\eta$, this yields a positive key rate for $\eta > 91.1\%$. Considering again the P\&M case with $\eta_B^* \to 1$, we obtain the lower bound $\eta_A^- > 83.3\%$. These are the figures appearing in Table 1 in the main text. In both cases, but especially the P\&M case, using our post-selection technique yields substantially better efficiency thresholds than the non-post-selected protocols.
We note that somewhat lower efficiency thresholds for DI-QKD ($\eta > 92.3\%$ and $\eta > 88.9\%$ for the entanglement-based schemes, with and without post-selection, resp.) can be obtained using the analysis of Ref.~\cite{DIQKD_NJP_SM}. However, this analysis shows security only against collective attacks (which includes, in particular, the assumption that the devices are memoryless).

In the main text we presented only the post-selected version of 1sDI-QKD. However one can also consider a non-post-selected version of that, in which Alice and Bob try to extract a secret key directly from the non-post-selected data ${\boldsymbol A}_1, {\boldsymbol B}_1$ (rather than from ${\boldsymbol A}_1^{\rm ps},{\boldsymbol B}_1^{\rm ps}$; note however that we still post-select on a detection by Bob). In this case the key rate per detection by Bob is bounded from below by $1 - h(Q_1) - h(Q_2)$ for security against coherent attacks. Here for a typical implementation the bit error rates are both given by $Q_i = (1-V\eta_A)/2$, and with $V=1$, we get a positive rate for $\eta_A > 78.0\%$. Recall that when Alice and Bob use the  post-selected data, the threshold is $\eta_A > 65.9\%$, so the improvement is even more dramatic than in the fully DI case.


\begin{thebibliography}{99}
\bibitem{footnote_finite_key_effects} The approximate (in)equalities in the derivation of Eq.~(\ref{eq_l_1sDI_approx}) account for additional correcting terms that appear in the equations in the case of finite key lengths. One typically finds that for raw key lengths longer than $10^4 - 10^5$ bits, significant fractions of the asymptotic secret key rates are obtained~\cite{TomaLimGiRenner}.
\bibitem{endnote:2}{To avoid any confusion, we note that here we have reversed the roles of Alice and Bob compared to that in Refs.~\cite{TR11,TomaLimGiRenner}, in order to be consistent with earlier work on steering \cite{steering_PRL,CJWR09}, in which Alice's detectors are not trusted, while Bob's are.}
\end{thebibliography}
\end{document}